\RequirePackage{fix-cm}

\documentclass[twocolumn,epjc3]{svjour3}

\RequirePackage{graphicx}

\usepackage{amsmath,graphicx,color,hyperref}

\def\bra{\langle}
\def\ket{\rangle}
\def\bbra{\langle\!\langle}
\def\kket{\rangle\!\rangle}

\newcommand{\trento}{T$\mathrel{\protect\raisebox{-2.1pt}{R}}$ENTo}
\newcommand{\pt}{$[p_t]$}
\newcommand{\By}{$\bra B_y \ket$}

\usepackage[dvipsnames]{xcolor}

\journalname{Eur. Phys. J. A}

\begin{document}

 \title{Manipulating strong electromagnetic fields with the average transverse momentum of relativistic nuclear collisions}

 \author{Giuliano Giacalone\thanksref{addr1} \and Chun Shen\thanksref{addr2,addr3}}
 

\institute{Institut f\"ur Theoretische Physik, Universit\"at Heidelberg,
 Philosophenweg 16, 69120 Heidelberg, Germany \label{addr1}
           \and
           Department of Physics and Astronomy, Wayne State University, Detroit, Michigan 48201, USA \label{addr2}
           \and
            RIKEN BNL Research Center, Brookhaven National Laboratory, Upton, NY 11973, USA \label{addr3}
            }

\date{}

\maketitle

\begin{abstract}
We show that an event-shape engineering based on the mean transverse momentum of charged hadrons, $[p_t]$, provides an optimal handle on the strength of the magnetic field created in central heavy-ion collisions at high energy. This is established through quantitative evaluations of the correlation existing between the event-by-event magnetic field produced by the spectator protons in 5.02 TeV Pb+Pb collisions and the event-by-event $[p_t]$ at a given collision centrality. We argue that the event selection based on $[p_t]$ provides a better handle on the magnetic field than the more traditional selection based on the event ellipticities. Advantages brought by this new method for the experimental search of the chiral magnetic effect are discussed.
\end{abstract}

\section{Introduction}

The interaction of two heavy nuclei at relativistic energy gives rise to a short-lived electromagnetic (EM) field of gigantic strength. The charge carried by the protons lying at the edges of the colliding ions, and that fly along the beam pipe without undergoing any interactions (the so-called \textit{spectator} protons), engenders in particular a magnetic field over the transverse plane. Since the two nuclei move along opposite directions, the field lines coming from their spectators sum up coherently over the interaction region. The resulting magnetic field is the strongest ever produced in a laboratory, of order $e|\vec B|\approx m_{\pi}^2$ \cite{Skokov:2009qp} (or $10^{14}$~T in SI units).

A vast literature is devoted to studying the phenomenological consequences of the strong coherent field produced in heavy-ion collisions. Detecting signatures of the EM field would represent an important new result in nuclear physics, and would permit to test the fascinating prediction that the a strong magnetic field coupled with the ultra hot-and-dense quark-gluon plasma created in high-energy nuclear collisions may lead to observable effects due to local parity violation in the strong sector~\cite{Li:2020dwr}, the most notorious of which is the chiral magnetic effect (CME) \cite{Fukushima:2008xe}. The observable that seems intrinsically connected with the manifestation of the strong $\vec B$ field in heavy-ion collisions is the dipolar flow of charged hadrons, $v_1$, as recently discussed in Ref.~\cite{Oliva:2020mfr}. The CME is indeed a charge-dependent $v_1$, driven by an electric current flowing along the $\vec B$ field direction in presence of local parity violation \cite{Kharzeev:2004ey}. The rapidity profiles of the charge-dependent $v_1$ are further expected to be sensitive to the EM field \cite{Gursoy:2018yai,Oliva:2019kin}, especially for heavy hadrons, such as $D$ mesons~\cite{Das:2016cwd,Chatterjee:2017ahy,Chatterjee:2018lsx}, that are more sensitive to the early-time dynamics of the collision process where the $\vec B$ field is the strongest.

Despite the great interest from the theoretical community, and several experimental explorations at both the Relativistic Heavy Ion Collider (RHIC) \cite{Abelev:2009ac,Abelev:2009ad,Adamczyk:2013kcb,Adamczyk:2014mzf,STAR:2019xzd,Adam:2019wnk,STAR:2020crk,Adam:2020zsu} and the Large Hadron Collider (LHC) \cite{Abelev:2012pa,Khachatryan:2016got,Sirunyan:2017quh,Acharya:2019ijj,Acharya:2020rlz,Sirunyan:2020dop}, a smoking-gun of the manifestation of the strong $\vec B$ field in heavy-ion collisions is still missing. The problem is arguably that the signatures so far predicted by the theoretical models are very feeble, which makes it difficult to draw any definite conclusions from the experimental data. The predicted probes of the EM field are in general charge-dependent flow coefficients or other observables that in high-energy collisions are dominated by the hydrodynamic expansion of the quark-gluon plasma \cite{Tuchin:2013ie}. These probes only receive small perturbative corrections from the short-lived EM field \cite{Voronyuk:2011jd,Das:2017qfi,Roy:2017yvg,Stewart:2017zsu,Inghirami:2019mkc}. As a consequence, one typically ends up thus in the uncomfortable situation where the signal-to-background ratio for the $\vec B$ field signatures is of order 1\%, or less. 

In Ref.~\cite{Giacalone:2020oao}, a novel method to enhance and thus potentially observe the manifestations of the strong $\vec B$ field is proposed. The idea is to build correlations between the relevant charge-dependent observables, such as $v_1$, and the average transverse momentum of all charged hadrons, \pt{} $\equiv (\sum_i^{N_\mathrm{ch}} p_t^i)/(\sum_i^{N_\mathrm{ch}} 1)$, detected in a given centrality class. The point is that, at a given centrality (or multiplicity), \pt{} is in a strong correlation with the number of spectator nucleons, especially for central collisions. By varying the value of \pt{}, one can thus effectively turn up and down the number of spectators, potentially leading to sizable variations in the $\vec B$ field that they induce.

In this paper, we validate this idea by establishing on quantitative grounds the connection between the event-by-event \pt{} and event-by-event magnetic fields at a given centrality. We perform simulations of the initial state of Pb+Pb collisions at top Large Hadron Collider (LHC) energy, and, by use of appropriate estimators of the final-state quantities, we assess to which extent the $\vec B$ field and \pt{} are correlated at a given centrality, and what advantages the experimental analysis of such a correlation brings with respect to more traditional methods.

 This article is organized as follows. In Sec.~\ref{sec:2}, we describe the setup of our simulations, i.e., how we model the collision process, how the magnetic field is calculated in each event, and what properties of the initial state are used in the subsequent evaluations. In Sec.~\ref{sec:3}, after explaining the physical meaning of correlating observables with \pt{}, we review the idea of Ref.~\cite{Giacalone:2020oao}, and present our main result, i.e., an estimate of the correlation between \pt{} and the $\vec B$ field in central heavy-ion collisions, whose implications are then discussed in Sec.~\ref{sec:4}. We conclude with Sec.~\ref{sec:5}, where we discuss the new directions of investigations, both theoretical and experimental, opened by our findings.

\section{Collision model, EM field, and initial state}
\label{sec:2}

\subsection{Collision model}

We employ a Monte Carlo Glauber-type description of Pb+Pb collisions at top LHC energy, following the \trento{} model of initial conditions \cite{Moreland:2014oya}. The colliding nuclei are treated as batches of 208 nucleons sampled independently from a common single-particle density, namely, a two-parameter Fermi distribution:
\begin{equation}
\rho(r)\propto 1 / \bigl[ \exp((r-R)/a) \bigr],   
\end{equation}
where $R=6.62$~fm is the half-density radius, and $a=0.55$~fm is the skin thickness \cite{DeJager:1987qc}. A minimum distance $d_\mathrm{min} = 0.8$ fm between the sampled nucleons is imposed. We do not make any distinction between neutron and proton densities, although they are not identical \cite{Abrahamyan:2012gp,Tarbert:2013jze}. We assume in practice that the individual nucleon has a probability $Z/A$ to be a proton, and make clear in the following that our results do not depend on this approximation.

The two nuclei overlap at some impact parameter, $b$, and nucleon-nucleon  (NN) collisions take place depending on the pair distance in the transverse plane and the NN inelastic cross section at the given collision energy. Nucleons are thus labeled as \textit{participants} if they undergo at least one interaction with a nucleon from the other nucleus, and \textit{spectators}, if they fly unscattered away from the interaction region along the beam direction. These spectators are especially important in our analysis, as we shall evaluate the magnetic fields which they induce over the collision area. We denote the number of spectator nucleons in a collision event by $N_s$. 

In one nucleus, each participant nucleon is turned into a transverse density of participant matter by centering on top of its location a two-dimensional Gaussian profile of width $w=0.8$~fm, and random normalization sampled from a gamma distribution of unit mean and unit variance. These parameters for the gamma fluctuations, as well as the choice of the width $w$ are motivated by recent studies that infer high-probability \trento{} model parameters from comparisons of hydrodynamic simulations with Pb+Pb data \cite{Bernhard:2019bmu,Everett:2020yty,Everett:2020xug}. Note that $w=0.8$~fm implies rather fat nucleons, leading to smooth profiles of energy density in the transverse plane (as explicitly shown later on in Fig.~\ref{fig:5}), a feature which turns useful in our evaluations of the magnetic field induced by the spectator protons, as we clarify below.

\subsection{Energy deposition}
The superimposition of the Gaussian participant nucleon densities gives the total density of participant matter in a given nucleus, say $A$, which we dub $T_A({\bf x})$, where ${\bf x}$ labels a coordinate in the transverse plane. The energy density in units GeV/fm$^2$ deposited in the midrapidity slice (on which our analysis is focused) right after the collision ($\tau=0^+$) of nucleus $A$ against nucleus $B$ is finally obtained as:
\begin{equation}
\label{eq:edens}
    e({\bf x},0^+) = N_e \sqrt{T_A({\bf x})T_B({\bf x})},
\end{equation}
where we use $N_e=18$~GeV \cite{Bernhard:2019bmu}. The total initial energy at midrapidity is thus equal to:
\begin{equation}
\label{eq:E}
    E = \int d^2{\bf x} ~e({\bf x},0^+).
\end{equation}

\begin{figure}[t]
    \centering
    \includegraphics[width=.8\linewidth]{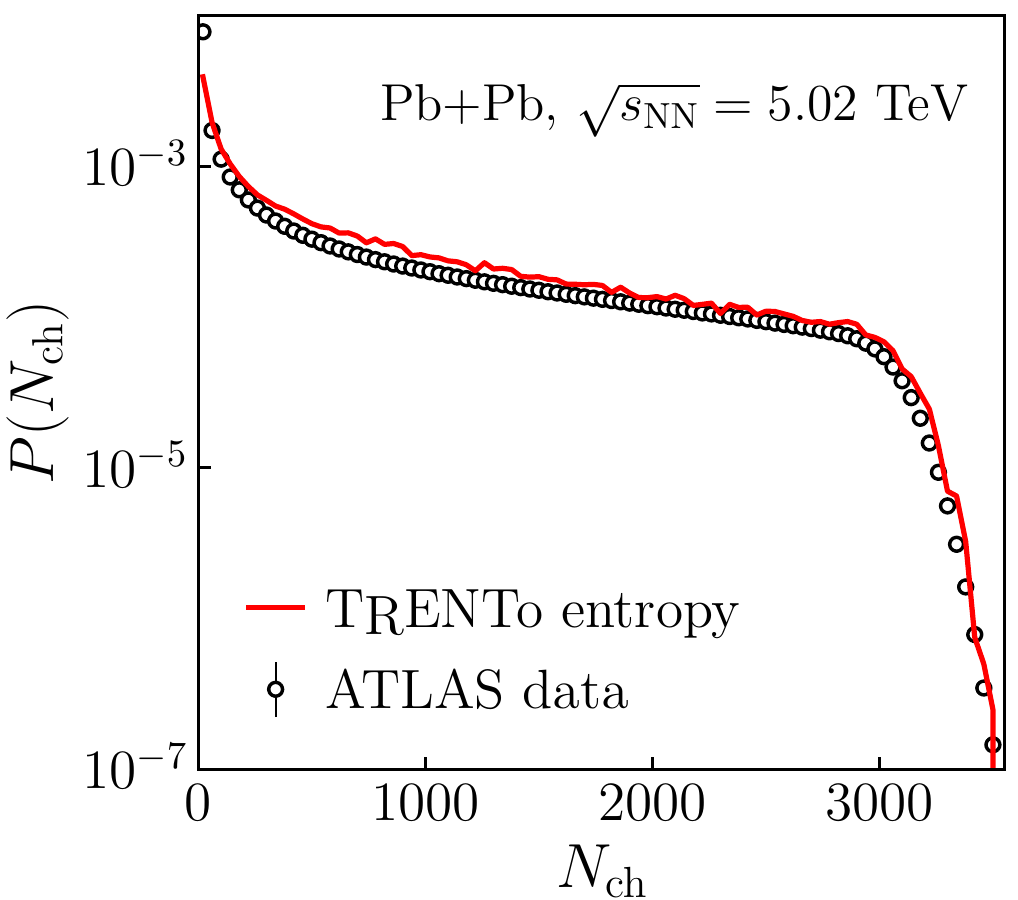}
    \caption{Symbols: ATLAS data \cite{Aaboud:2019sma} on the minimum bias distribution of charged multiplicity, $N_{\rm ch}$, measured in the pseudorapidity acceptance $|\eta|<2.5$, in 5.02 TeV Pb+Pb collisions. Line: rescaled probability distribution of the event-by-event \trento{} entropy, $S$, in $10^5$ simulated events.}
    \label{fig:1}
\end{figure}

We run a sample of $10^5$ minimum bias Pb+Pb collisions. To sort our events into centrality classes, we consider that the initial entropy of the system is proportional to the multiplicity in the final state of the collisions\footnote{That is, we ignore the viscous entropy production during the hydrodynamic phase as well as Poisson fluctuations in the particlization stage.}, which is the quantity employed experimentally to define the centrality. We assume an instantaneous thermalization of the system and a conformal equation of state. The entropy density of the system in this simple picture is given by:
\begin{equation}
    s({\bf x}) \propto e({\bf x})^{4/3},
\end{equation}
so that the total entropy at midrapidity is:
\begin{equation}
\label{eq:S}
    S = \int d^2{\bf x} ~s({\bf x}).
\end{equation}
Upon an appropriate rescaling\footnote{In practice, we use the Bayesian inversion of Ref.~\cite{Das:2017ned} to extract from the experimental data the average value of $N_{\rm ch}$ in collisions at zero impact parameter. We find $\bra N_{\rm ch} \ket (b=0) = 3110$. This value is then compared to the value of $\bra S \ket (b=0)$ in the \trento{} model to extract the proportionality factor used to draw the histogram in Fig.~\ref{fig:1}.}, the histogram of $S$ obtained in the \trento{} model is compared in Fig.~\ref{fig:1} to the experimental distribution of the the raw charged multiplicity, $N_{\rm ch}$, measured by the ATLAS collaboration \cite{Aaboud:2019sma}. The simple \trento{} prescription captures with an excellent accuracy the shape of the experimental histogram, which justifies the use of $S$ as a variable to sort our events into centrality classes. A little mismatch is observed at the level of the global normalization of the two histograms, due to the imperfect description of the \trento{} model of the very high values of probability observed experimentally in the most peripheral bin, $N_{\rm ch}\in[0,40]$.

\subsection{Magnetic field}

On an event-by-event basis, we evaluate the magnetic field induced over the transverse plane by the spectator protons (by assuming, as stated above, that a $Z/A$ fraction of spectators are protons). We calculate the magnetic field with the tools developed in Refs.~\cite{Gursoy:2014aka,Gursoy:2018yai}, which we modify to include event-by-event fluctuations in the position of the spectators. For a given spectator, we evaluate the magnetic field induced in the medium formed in the collision, under the assumption that this medium has an electric conductivity $\sigma=0.023$~fm$^{-1}$ which is the same at all space-time points. This allows one to derive a semi-analytical solution for the space-time evolution of the EM field. We are given a spectator proton located at position ${\bf x}'=(x',\phi')$ in the transverse plane, and moving along either the positive or the negative space-time rapidity ($\eta_s$) direction with beam rapidity $Y$ for projectile and target spectators, respectively. At a given point ${\bf x}=(x,\phi)$ in the transverse plane, this moving charge induces a magnetic field along the $y$ direction, i.e., the direction orthogonal to the impact parameter, given by:
\begin{align}
\nonumber    B_y (\tau, \eta_s, x, \phi ) &= \alpha_e \sinh(Y) (x\cos \phi - x'\cos \phi') \\
    & \times \frac{\sigma |\sinh(Y)| \sqrt{\Delta} / 2 + 1 }{\Delta^{3/2}} e^\mathcal{A},
\end{align}
where $\alpha_e$ is the electromagnetic coupling, while:
\begin{align}
    \nonumber\mathcal{A} &= \frac{\sigma}{2} \bigl [ \tau \sinh(Y)\sinh(Y-\eta_s) - |\sinh(Y)|\sqrt{\Delta} \bigr], \\
    \Delta &= \tau^2 \sinh^2(Y-\eta_s) + x^2 + (x')^2 -2 x x' \cos(\phi-\phi').
\end{align}
Similar equations can be used to evaluate as well the magnetic field along the direction orthogonal to $y$, $B_x$, and the components of the electric field, $E_x$ and $E_y$, which we shall not use in our analysis. We neglect any influence on $B_y$ coming from the participant nucleons, whose space-time dynamics is more complicated due to their locations inside the medium. The inclusion of the participants should however have a minor influence on our discussion. Participant nucleons lead to strong local fluctuations of magnitude of $\vec B$ \cite{Bzdak:2011yy,Deng:2012pc}, but as their field lines close inside the medium, they should not impact significantly to the net field along the $y$ direction.

 The superimposition of the $\vec B$ fields coming from the spectators has to be evaluated over the region where the medium lies. While doing so, it is important to avoid including contributions from regions in the transverse plane where the energy density of the medium is negligible, as these contributions would lead to an incorrect depletion of the strength of $B_y$. This involves some degree of arbitrariness in the definition of the area over which $\vec B$ is evaluated. Our choice is that of calculating the event-by-event magnetic produced by the spectator nucleons as an average of $\vec B$ weighted by the energy density of the created medium, that is:
\begin{equation}
\label{eq:Byave}
    \bra \vec{B} \ket = \frac{1}{E} \int d^2{\bf x} ~\vec{B}({\bf x}) e({\bf x}).
\end{equation}
This implies an artificial enhancement of the $\vec B$ field in the regions of overdensity. However, as mentioned above the \trento{} model used in our application yields event-to-event energy density profiles that are rather smooth ($\delta e({\bf x})/e ({\bf x})\sim3$ at maximum, as also observed in Fig.~\ref{fig:5}). The $\sqrt{T_AT_B}$ in Eq.~(\ref{eq:edens}) prescription further ensures that the energy density profiles fall off rapidly away from the center of the fireball. The energy-density weight in Eq.~(\ref{eq:Byave}) seems thus a very reasonable choice.

\begin{figure}[t]
    \centering
    \includegraphics[width=.8\linewidth]{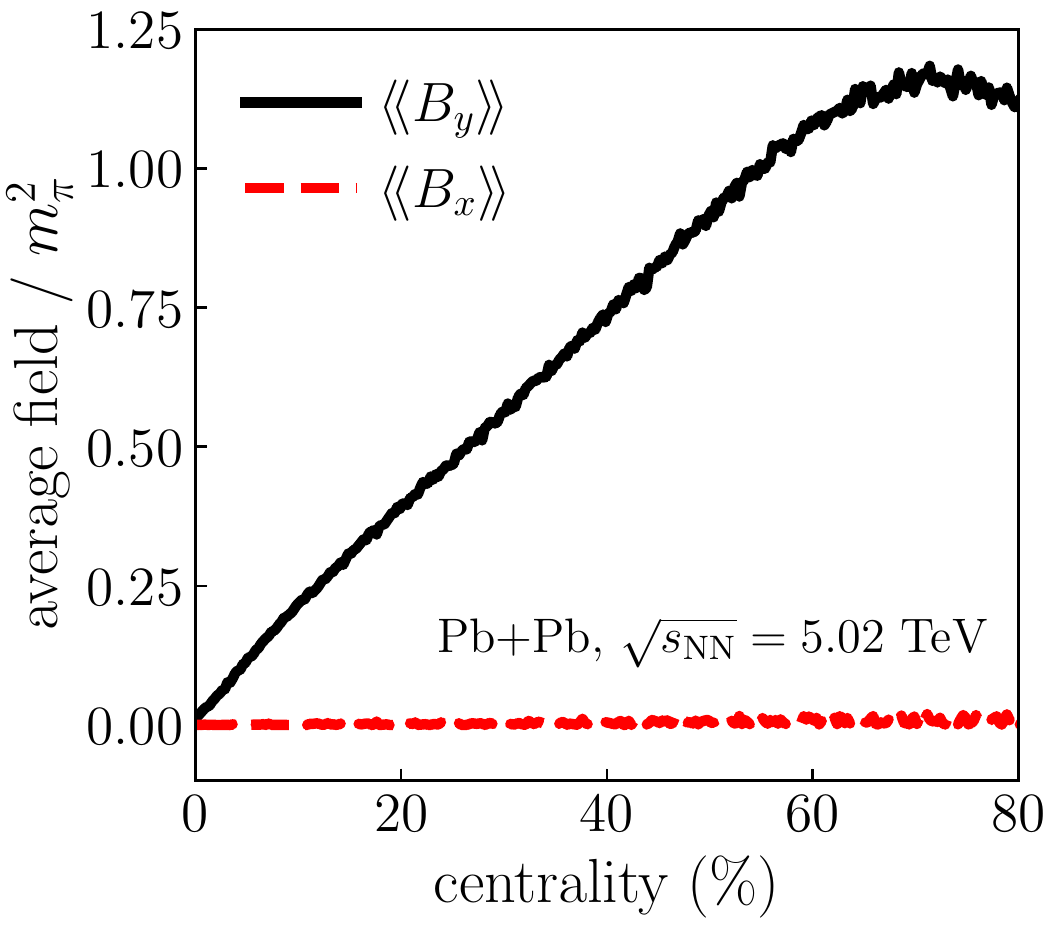}
    \caption{Average of the event-by-event average components of the magnetic field, as given by Eq.~(\ref{eq:Byave}), induced by the spectator protons in the midrapidity slice, at $\tau=0^=$ in 5.02 TeV Pb+Pb collisions, as a function of the centrality percentile. Solid line: $\bbra B_y \kket$. Dashed line: $\bbra B_x \kket$.}
    \label{fig:2}
\end{figure}

The average magnetic field of the spectator protons is shown as a function of collision centrality in Fig.~\ref{fig:2}. The event-average $\bbra B_x \kket$ is identically zero, as expected from symmetry arguments. The average $\bbra B_y \kket$ does on the other hand grow linearly with the centrality percentile, driven by the increase of the number of spectators, $N_s$, with the collision impact parameter. It reaches a maximum of about $1~m_\pi^2$ in peripheral collisions.

\subsection{Estimators of the average transverse momentum}

The goal of this paper is to evaluate the statistical correlation between $\bra B_y \ket$ and the average transverse momentum of charged hadrons at midrapidity:
\begin{equation}
    [p_t] = \frac{1}{N} \int d^2{\bf p}_t \frac{dN}{d^2{\bf p}_t} |{\bf p}_t|.
\end{equation}
where $N$ is the charged multiplicity, and $dN/d^2{\bf p}_t$ is the hadron spectrum. In this paper, we do not perform full hydrodynamic simulations, but simply estimate the event-by-event \pt{} from the initial state of the collisions. Powerful initial-state predictors for \pt{} (at fixed multiplicity) have been recently discussed in the context of Pb+Pb collisions \cite{Bozek:2020drh,Gardim:2020sma,Schenke:2020uqq,Giacalone:2020dln,Bozek:2021zim}. The average transverse momentum of a heavy-ion collision is, at a given multiplicity, in a tight correlation with the thermodynamic properties of the system \cite{Gardim:2019xjs}, in particular the system size, to be defined below, and the total energy per unit rapidity, i.e., $E$ defined by Eq.~(\ref{eq:E}). 

We employ, thus, two different estimators for the average transverse momentum. The first exploits the strong correlation between \pt{} and $E$. The value of \pt{} provides a measure of the energy per particle in the final state. In the initial state, this corresponds roughly to the energy of the system divided by its entropy, i.e.,
\begin{equation}
    [p_t] \propto E/S,
\end{equation}
where $E$ and $S$ are defined event-to-event by Eq.~(\ref{eq:E}) and (\ref{eq:S}), respectively. The second estimator that we consider exploits on the other hand the tight negative correlation between \pt{} and the system size. We predict thus \pt{} by means of an optimal measure of the (inverse) system size \cite{Schenke:2020uqq}:
\begin{equation}
    [p_t] \propto S/A_e,
\end{equation}
where $A_e$ is the elliptical area:
\begin{equation}
    A_e = \pi R^2 \sqrt{1-\varepsilon_2^2},
\end{equation}
with $R^2$ being the mean squared radius of the energy density profile:
\begin{equation}
    R^2 = \frac{1}{E} \int d^2{\bf x}~ e({\bf x},0^+)~ |{\bf x}|^2,
\end{equation}
and $\varepsilon_2$ its eccentricity:
\begin{equation}
    \varepsilon_2 =  \frac{1}{E R^2} \left | - \int d^2{\bf x} ~e({\bf x},0^+)~ |{\bf x}|^2 e^{i2\phi} \right |.
\end{equation}

The eccentricity will be further used as a predictor for the event-by-event elliptic flow, $V_2$, of the collision, i.e., the second Fourier harmonic of the azimuthal spectrum:
\begin{equation}
    V_2 = \frac{1}{N} \int d^2{\bf p}_t \frac{dN}{d^2{\bf p}_t} e^{i2p_\phi},
\end{equation}
In a given centrality class the magnitude of elliptic flow, $v_2=|V_2|$, is almost in one-to-one correspondence with $\varepsilon_2$, i.e., one can consider
\begin{equation}
    v_2 \propto \varepsilon_2
\end{equation}
at fixed multiplicity. This is an excellent approximation in central heavy-ion collisions.

\subsection{Pearson coefficients}

In this manuscript we present mainly results for the statistical correlation between the quantities discussed earlier in this section. We quantify such correlation by means of Pearson correlation coefficients, a standard tool in the phenomenology of heavy-ion collisions. Given two observables, $o_1$ and $o_2$, their statistical correlation is defined by:
\begin{equation}
    \rho(o_1,o_2) = \frac{\bra \delta o_1 \delta o_2 \ket}{\sqrt{\bra (\delta o_1)^2 \ket}\sqrt{\bra (\delta o_2)^2)}},
\end{equation}
where the brackets indicate an average over events in a given centrality class, and we have introduced:
\begin{equation}
\label{eq:Pearson}
    \delta o = o - \bra o \ket.
\end{equation}
When $\rho(o_1,o_2)=1$ (or -1), the two observables are perfectly correlated (or anticorrelated). A significant correlation between observables corresponds typically to $|\rho|\geq 0.1$.

\section{Average transverse momentum as a handle on the early-time magnetic field}
\label{sec:3}

\subsection{Correlating observables with \pt{}: a new tool for heavy-ion collisions}

The possibility of analyzing correlations between the event-by-event \pt{} and observables such as the flow coefficients has not been considered in the phenomenology of heavy-ion collision until the end of 2015, when the correlation between \pt{} and $v_2$ as a byproduct of the principal component analysis of Ref.~\cite{Mazeliauskas:2015efa}. Shortly after, the same correlation was formulated by Bo\.zek \cite{Bozek:2016yoj} through a Pearson correlation coefficient, amenable in a straightforward manner to experimental investigations. The measurement of this coefficient was published by the ATLAS collaboration three years later, in 2019 \cite{Aad:2019fgl}. 

From the discussion of the previous section, we evince that evaluating the correlation between \pt{} and some observable $o$ at fixed multiplicity is tantamount to answering the following practical question: how does $o$ vary if one performs an isentropic transformation of the underlying medium which increases the temperature and reduces the volume? An example of such a transformation is illustrated here in Fig.~\ref{fig:5}. The figure shows two profiles of energy density created in two central Pb+Pb collisions that present the same multiplicity (i.e. entropy), but values of \pt{} (estimated through $E/S$) that are largely different. One can see that the event at high \pt{} (panel on the right) corresponds to a medium that is more compact and more dense. This is indeed the physical meaning and implication of modifying \pt{} at a given centrality.

Recently, correlations based on \pt{} have attracted great attention in the community with the realization that they allow one to magnify the manifestations of remarkable nuclear phenomena that would otherwise be very difficult to identify with more conventional tools. Two such phenomena have so far been worked out in the literature, namely, $i)$ the fact that correlating $v_2$ with \pt{} allows one to obtain spectacular signatures of the deformation of the colliding ions \cite{Giacalone:2019pca}, in particular, to gather evidence of polarized body-body collisions of $^{238}$U nuclei (as recently verified experimentally \cite{jia}); $ii)$  the fact that, for collisions at small multiplicities, the sign of the correlation between \pt{} and $\varepsilon_2$ is different than the sign of the correlation between \pt{} and the eccentricity of the initial condition of the system in momentum space \cite{Giacalone:2020byk}. Thanks to this sign difference, it should be possible to use the correlation between $v_2$ and \pt{} as a probe of this initial-state momentum anisotropy, which in turn would probe correlations of gluon fields, predicted in particular by the color glass condensate theory of high-energy QCD \cite{Altinoluk:2020wpf}, in the earliest stages of the collision process.

In this manuscript, we add another item to this list. From the results presented in this section, we shall argue that by correlating appropriate observables with \pt{} one can enhance the phenomenological manifestation of the strong $\vec B$ field created in heavy-ion collisions.

\subsection{The idea: spectator nucleons and \pt}

Our analysis is based on the idea introduced in Ref.~\cite{Giacalone:2020oao}. The key point is that the variation of system size induced by a variation of \pt{} at fixed multiplicity is accompanied by a rather strong variation of the number of spectator (or participant) nucleons. This is shown explicitly in Ref.~\cite{Giacalone:2020oao}, where for central Pb+Pb collisions it is reported an increase in $N_s$ of a large factor by moving from low-\pt{} to large-\pt{} collisions.

\begin{figure}[t]
    \centering
    \includegraphics[width=.81\linewidth]{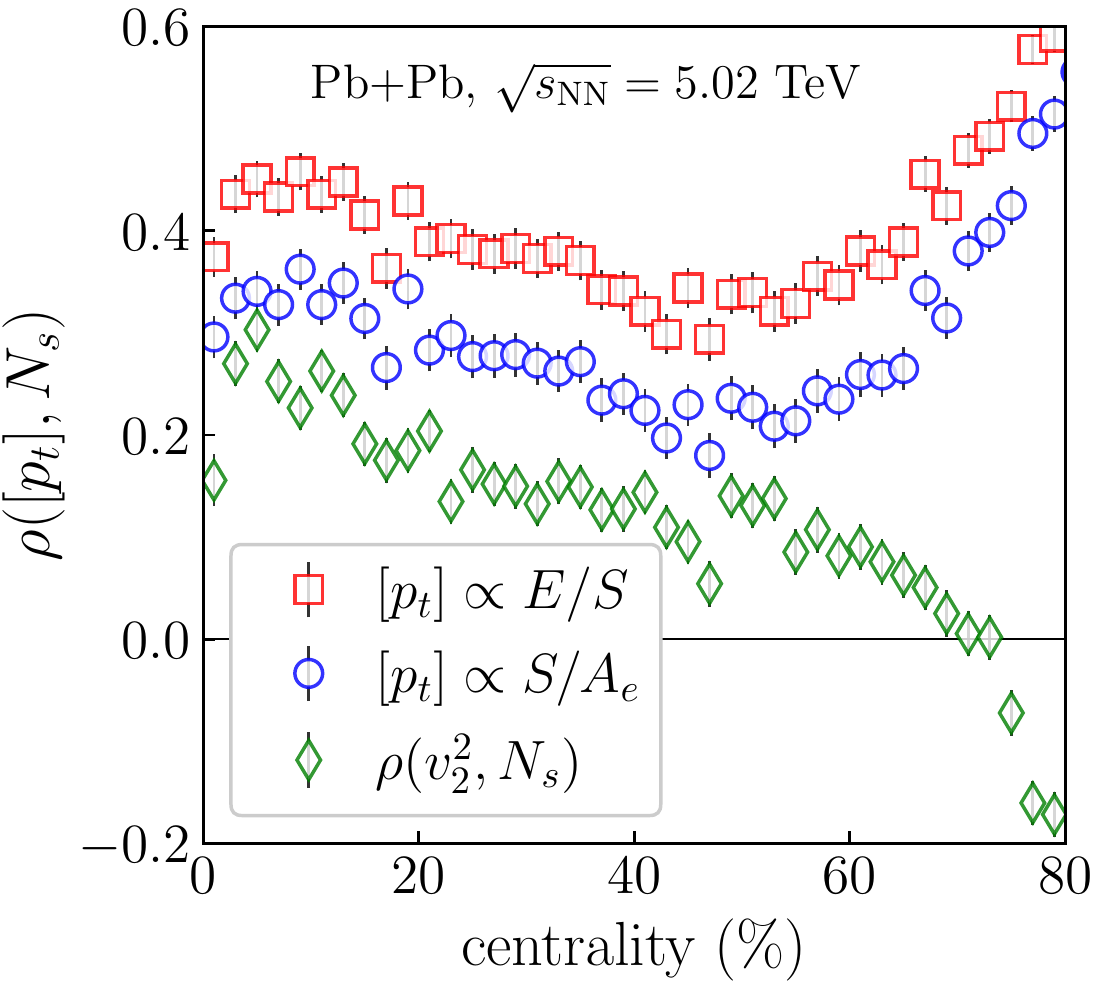}
    \caption{Correlation between \pt{} and $N_s$, $\rho([p_t],N_s)$ as a function of centrality percentile in 5.02 TeV Pb+Pb collisions. Squares: $[p_t] \propto E/S$. Circles: $[p_t]\propto S/A_e$. The diamonds show instead the correlation $\rho(v_2^2, N_s)$.}
    \label{fig:3}
\end{figure}

We go now a little beyond that simple analysis and evaluate the Pearson correlation between \pt{} and $N_s$, $\rho([p_t],N_s)$, as defined by Eq.~(\ref{eq:Pearson}), across the centrality percentile. To ensure that the correlator is evaluated at fixed multiplicity, we first evaluate it into ultra-narrow centrality bins of width 0.25\%, which we then recombine into bins of width 2\%. The results are shown in Fig.~\ref{fig:3}, where we consider both $[p_t]\propto E/S$ (red squares), and $[p_t]\propto S/A_e$ (blue circles).  We observe, for both predictors, a significant correlation between $N_s$ and $\bra p_t \ket$. The correlation is stronger in central collisions than in mid-peripheral collisions, in agreement with the findings of Ref.~\cite{Giacalone:2020oao}. 

In Figure~\ref{fig:3} we show as well, as green diamonds, $\rho(v_2^2, N_s)$, obtained by considering $v_2\propto \varepsilon_2$. This curve is systematically (and significantly) below the other ones, which suggests that an event-shape engineering based on \pt{} provides a better handle on $N_s$ than a selection based of the event-by-event $v_2$, often considered in the literature \cite{Dobrin:2012zx,Beraudo:2018tpr}. Our conclusion is that \pt{} is indeed the final-state observable presenting the strongest correlation with $N_s$ at a given multiplicity. We shall come back to this point in Sec.~\ref{sec:4}.

\subsection{Correlation of $\bra B_y \ket$ with \pt}

Having explained the physical picture motivating the correlation of observables with \pt{}, and having crosschecked the proposal of Ref.~\cite{Giacalone:2020oao}, we can now move on to show our main results, involving the event-by-event net magnetic field, \By{}. 

\begin{figure}[t]
    \centering
    \includegraphics[width=.8\linewidth]{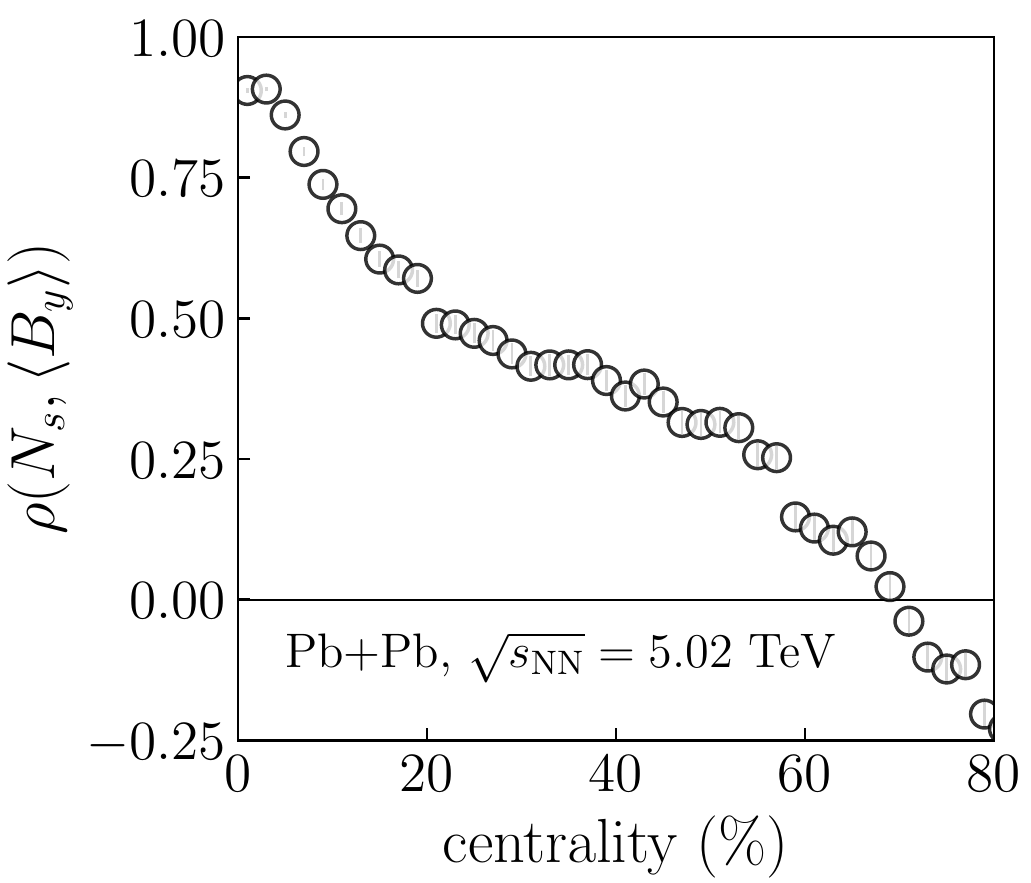}
    \caption{Correlation between the net spectator $\vec B$ field, \By{}, and the number of spectator nucleons, $N_s$, as a function of centrality percentile in 5.02 TeV Pb+Pb collisions. }
    \label{fig:4}
\end{figure}

To start with, we show in Fig.~\ref{fig:4} the correlation between the magnetic field and the spectator number, $\rho(N_s,\bra B_y \ket)$. The shape of the curve is quite interesting. In central collisions the correlation is almost perfect, and it gradually degrades as one moves towards peripheral centralities, which is not surprising. As one moves to large centralities, the fluctuations in the positions of the spectators relative to the center of the fireball become larger. As a consequence, having more spectators does not necessarily imply a larger coherent field over the interaction region. The correlation observed in Fig.~\ref{fig:4} becomes indeed negative at very large centralities, which is a striking phenomenon. Note also that the curve in Fig.~\ref{fig:4} does not simply follow from the curves shown in Fig.~\ref{fig:3}, confirming that a quantitative evaluation of the $\vec B$ field is necessary before drawing any conclusions about the applicability of the idea of Ref.~\cite{Giacalone:2020oao}.

\begin{figure*}[t]
    \centering
    \includegraphics[width=\linewidth]{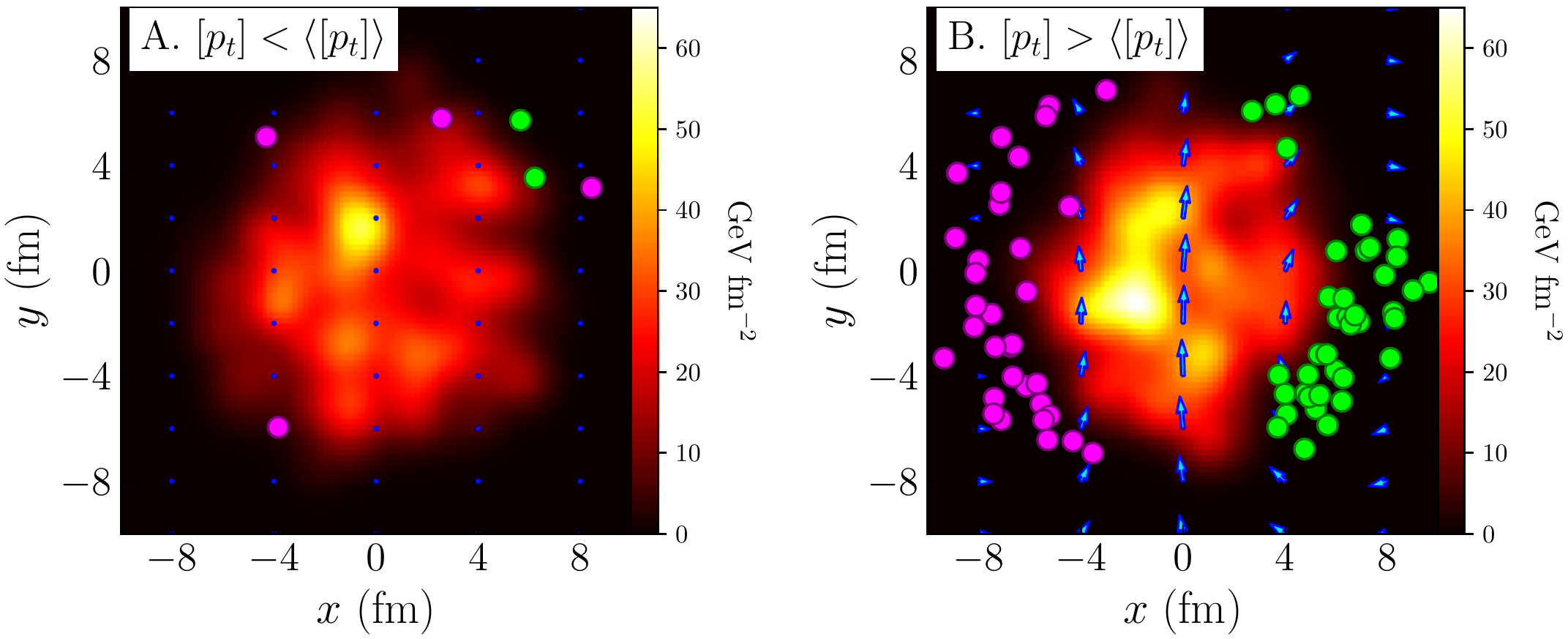}
    \caption{Left: initial energy density profile ($\tau=0^+$) of a central Pb+Pb collisions at top LHC energy corresponding to an abnormally low value of \pt{}. Right: same as in the left panel, but displaying a collision with an abnormally large value of \pt{} (at the same multiplicity). The magenta and green points indicate the spectators' transverse positions from projectile and target, respectively. The details of the event properties are listed in Tab.~\ref{tab:1}.}
    \label{fig:5}
\end{figure*}

In central collisions, then, one can safely expect that a variation of the number of spectator number is followed by a an analogous variation in \By{}. We check now whether such a variation can effectively be induced by means of \pt{}. It is useful to have an explicit illustration of the phenomenon we aim at describing, as shown in Fig.~\ref{fig:5}. We present two Pb+Pb events with nearly-identical high multiplicities (the events are taken from the 2-3\% centrality class) but significantly different values of \pt{}.\footnote{As shown in Table~\ref{tab:1}, the variation in the relative \pt{} from event A to event B is about 5\%, which may look tiny. But as a matter of fact, these two events are $\sim 5 \sigma$ apart from each other in the \pt{} distribution. In central Pb+Pb collisions, the measured standard deviation of relative \pt{} fluctuations is indeed smaller than 1\% \cite{Abelev:2014ckr}.} The properties of these two events are summarized in Table~\ref{tab:1}. We see in particular that transformation of the system from low \pt{} to large \pt{} changes in a dramatic way the number of spectator nucleons (shown as circles in the figure), $N_s$, which in this extreme example varies by over one order of magnitude.

We overlay now the energy density profiles with the lines of magnetic field induced by the spectator nucleons in each event. We can thus see the effect of varying \pt{} at play. For sake of visibility, in the figure we evaluate the field lines at proper time $\tau=0.4$ fm/$c$ after the collision, i.e., we let the spectators slide a little bit along the beam direction in order to remove from the midrapidity slice the uninteresting and strong field lines that surround their locations. In the two panels, the scale of the arrows is identical, therefore, the fact that the arrows are not visible in the event at low \pt{} is consistent with the fact that the event at high \pt{} presents a \By{} which is larger by more than one order of magnitude, as reported in Table~\ref{tab:1}. In this extreme example, then, the variation of \pt{} does effectively turn on the strong $\vec B$ field, which is the phenomenon we are after here.

\begin{table}[t]
\centering
\begin{tabular}{|c|c|c|}
\hline
event & A & B \\
\hline
$N_{\rm ch}$ & 2813  & 2791\\
$b$ (fm)  & 0.49 & 4.41 \\
$N_s$  & 6 & 72 \\
$[p_t]/\bra [p_t] \ket$  & 0.976 & 1.028  \\
$\bra B_y \ket/m_\pi^2$  &  0.013 & 0.151 \\
$R$ (fm)  & 4.53   &  4.03 \\
$\varepsilon_2$ & 0.073 & 0.153 \\
\hline
\end{tabular}
\caption{\label{tab:1} Properties of the events A and B shown in Fig.~\ref{fig:5}. All the quantities reported in the table are defined in Sec.~\ref{sec:2}.}
\end{table}

\begin{figure}[t]
    \centering
    \includegraphics[width=.85\linewidth]{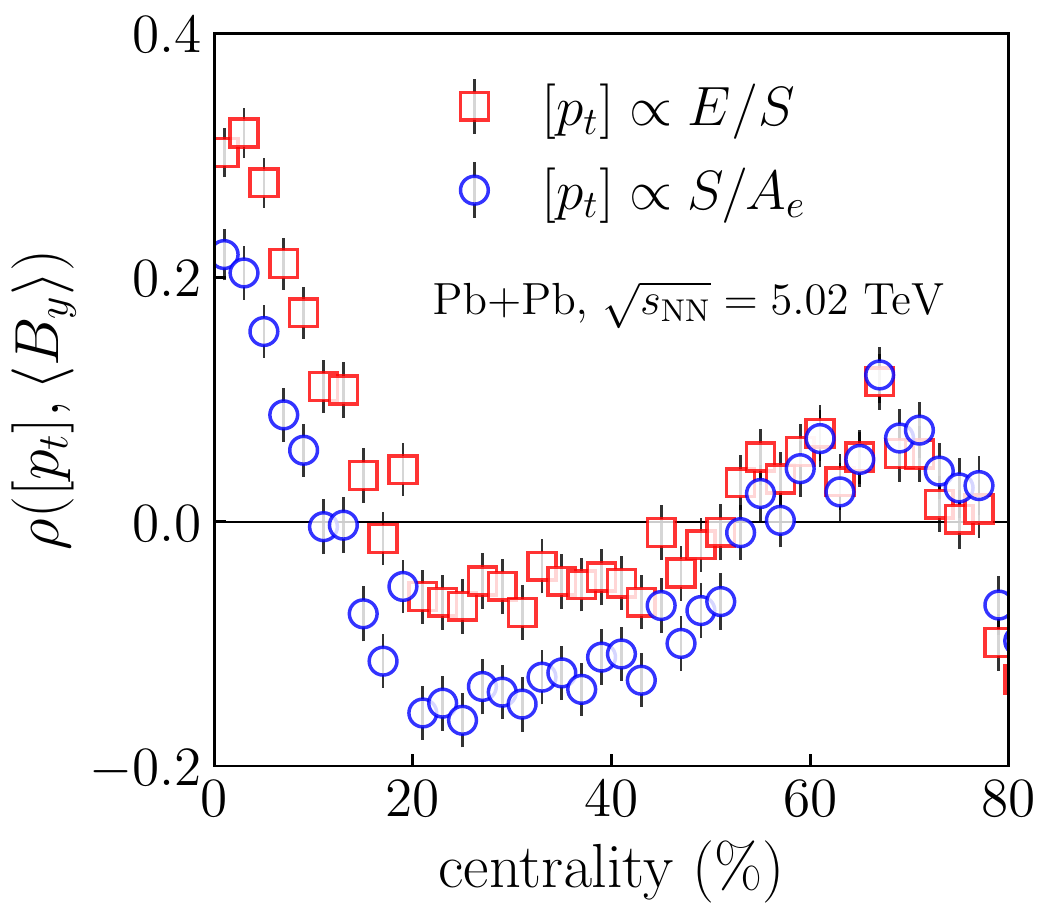}
    \caption{Correlation between \pt{} and \By{} in 5.02 TeV Pb+Pb collisions as a function of centrality percentile. Squares: $[p_t] \propto E/S$. Circles: $[p_t]\propto S/A_e$.}
    \label{fig:6}
\end{figure}

We quantify, then, the correlation between \pt{} and \By{} as a function of centrality percentile by evaluating the Pearson coefficient $\rho([p_t],\bra B_y \ket)$. Figure~\ref{fig:6} shows our main result. In the limit of central collisions, we obtain $\rho([p_t],\bra B_y \ket)\approx 0.3$, corresponding to a very sizable correlations. This provides conclusive indication that the idea proposed in Ref.~\cite{Giacalone:2020oao} is phenomenologically viable. 

The implications of this result are detailed in the next section.  An important comment is however in order. The result presented in Fig.~\ref{fig:6} does not come from a straightforward combination of the curves shown in Fig.~\ref{fig:3} and in Fig.~\ref{fig:4}. The correlator of Fig.~\ref{fig:6} does indeed drop quickly as a function of centrality percentile\footnote{We do not understand why our result becomes so significantly negative around 20\% centrality for the $S/A_e$ predictor. This requires further investigation with realistic hydrodynamic simulations. It is likely an artifact of the presence of $\varepsilon_2^2$ in the predictor.}, and become uninteresting, i.e., very close to zero, already at centralities of order 20\%. This is once again caused by the nontrivial interplay between the number of spectators and their transverse locations, which makes the manifestation of \By{} less straightforward to predict.

\section{An optimal event-shape engineering using \pt{}}
\label{sec:4}

Our main result, $\rho([p_t],\bra B_y \ket)\approx 0.3$ in central Pb+Pb collisions, indicates that by varying \pt{} at a given collision centrality one can thus turn up and down the strength of \By{}. This new method of inducing phenomenological manifestations of the strong magnetic field brings the following remarkable advantages.

\begin{figure}[t]
    \centering
    \includegraphics[width=.82\linewidth]{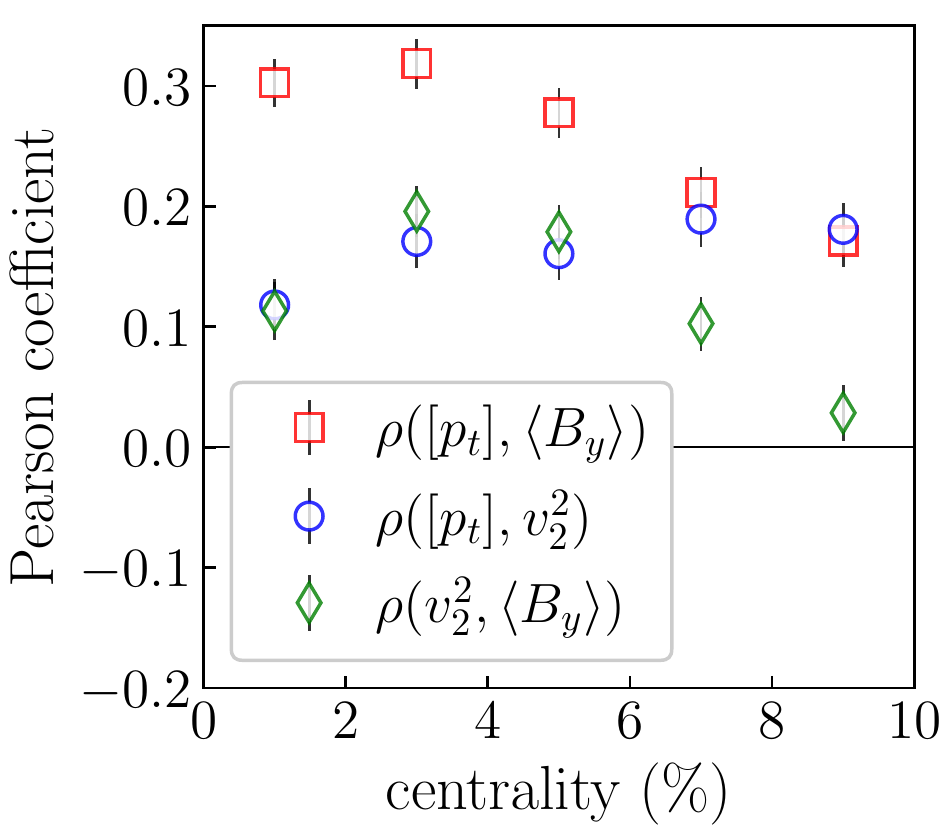}
    \caption{Pearson correlation coefficients, as defined by Eq.~(\ref{eq:Pearson}), in central Pb+Pb collisions at top LHC energy. We use $[p_t]\propto E/S$.  Squares: correlation between \By{} and \pt{}. Circles: correlation between \pt{} and $v_2^2$. Diamonds: correlation between $v_2^2$ and \By{}.}
    \label{fig:7}
\end{figure}

We focus on central collisions where the selection based on \pt{} is more effective. In Fig.~\ref{fig:7}, our result for $\rho([p_t], \bra B_y \ket)$ is plotted in the most central bins along with two new correlators, $\rho([p_t], v_2^2)$, and $\rho(v_2^2, \bra B_y \ket)$. The comparison between the curves shown Fig.~\ref{fig:7} has nontrivial implications:
\begin{itemize}
    \item We find $\rho([p_t],\bra B_y \ket)>\rho(v_2^2, \bra B_y \ket)$.  This hierarchy suggests that an increase of $[p_t]$ leads to a relative increase of $\bra B_y \ket$ which is significantly larger than the relative increase of \By{} that would be obtained through an increase of $v_2$. In other words, an event-shape engineering based on $v_2$ at a given centrality  is less effective than a selection of events based on $[p_t]$ if one wants to enhance (or suppress) the strength of the $\vec B$ field. The average transverse momentum stands, then, as the observable quantity (so far found) presenting the largest event-by-event correlation with the value of \By{} at fixed multiplicity.
    \item An increase in the value of \pt{} yields an increase of both \By{} and $v_2^2$. However, in Fig.~\ref{fig:7} we see that $\rho([p_t],\bra B_y \ket)>\rho([p_t], v_2^2)$, meaning that the induced relative increase of $\bra B_y \ket$ is larger than that of $v_2^2$. This feature is important for improving the signal-to-background ratio in the experimental search of the CME. We recall that the CME signal is a dipolar flow of hadrons with the same charge:
    \begin{equation}
        \bra \cos(\phi_1^\pm - \phi_2^\pm) \ket,
    \end{equation}
    which develops along the direction of the magnetic field, $B$, in presence of strong local parity violation. As soon as the collision impact parameter is large enough, a net \By{} emerges in the direction orthogonal to that of elliptic flow \cite{Bloczynski:2012en}. As originally realized by Voloshin \cite{Voloshin:2004vk}, the natural measure of the CME signal in collisions at large impact parameter is then:
    \begin{equation}
    \label{eq:gamma}
        \bra \cos(\phi_1^\pm + \phi_2^\pm - 2 \phi_3) \ket.
    \end{equation}
    Looking for a percent-level excess $v_1$ along the direction of $v_2$ on top of a strong hydrodynamic flow is difficult, and the main issue affecting the interpretation of data on the correlator in Eq.~(\ref{eq:gamma}). Our finding that $\rho([p_t],\bra B_y \ket)>\rho([p_t], v_2^2)$ means however that one can use $[p_t]$ to enhance \By{} more than $v_2^2$, leading to an improved isolation of the genuine CME signal. It is essentially the same idea motivating the study of isobaric collisions \cite{Skokov:2016yrj}, i.e. $^{96}$Ru+$^{96}$Ru and $^{96}$Zr+$^{96}$Zr collided at RHIC in 2018, where one aims at having identical $v_2^2$ but larger spectator field due to larger number of protons in $^{96}$Ru. Our method can be used in a single large system, and, thanks to the significant correlation observed in Fig.~\ref{fig:7}, it may be more effective.
\end{itemize}

To conclude, we emphasize an additional advantage brought by this new method, as pointed out at the end of Ref.~\cite{Giacalone:2020oao}. In the hunt for the CME, a major background contribution comes from so-called $1/N$ effects (where $N$ is the charged multiplicity) such as the conservation of global transverse momentum \cite{Wang:2009kd,Schlichting:2010qia,Bzdak:2012ia,Bozek:2017thv}. The correlations involving \pt{} (see e.g. the observables discussed in the next section) are however taken at fixed, large $N$, and should not be contaminated by any such effects, providing thus an optimal handle on the genuine manifestations of the strong EM field.

\section{Conclusion and outlook}

\label{sec:5}

In summary, our quantitative analysis gives clear indication that the proposal of Ref.~\cite{Giacalone:2020oao} is phenomenologically viable. Thanks to the strong correlation between \pt{} and $N_s$ at a given centrality, one can effectively use \pt{} to enhance (or reduce) the magnitude of the strong $\vec B$ field created in heavy-ion collisions, which we have evaluated on an event-by-event basis. The correlation found in central collisions is significant, $\rho([p_t],\bra B_y \ket)\approx0.3$.

We expect this finding to trigger new developments in the near future, both theoretical and experimental.

From the experimental side, we recommend the analysis of the correlation of the standard probes of the $\vec B$ field with the average transverse momentum. 
As suggested in Ref.~\cite{Giacalone:2020oao}, the most natural correlation is that between \pt{} and the charge-dependent $v_1$ in central heavy-ion collisions:
\begin{equation}
    \bigl \bra \delta [p_t] \cos(\phi_1^\pm - \phi_2^\pm) \bigl \ket.
\end{equation}
Another possibility is then to build a 4-particle correlation where one correlates the observable of Voloshin with \pt{}:
\begin{equation}
    \bigl \bra \delta [p_t] \cos(\phi_1^\pm + \phi_2^\pm - 2 \phi_3  )\bigl \ket,
\end{equation}
which would in turn involve the correlation between \pt{} and $v_2^2$. This observable may thus be particularly interesting in the comparison between central $^{197}$Au+$^{197}$Au collisions and central collisions of well-deformed $^{238}$U nuclei at RHIC, as these two colliding systems have opposite sign for the $\rho(v_2^2, [p_t])$ correlation due to the large quadrupole deformation of $^{238}$U \cite{Giacalone:2020awm,jia}.

From the theoretical side, the task is to provide a baseline from state-of-the-art hydrodynamic simulations for these charge-dependent correlations, which we plan to do in a follow-up work. This is relatively straightforward, and can be done, e.g., by extending the calculations of Ref.~\cite{Schenke:2019ruo} to include \pt{} dependent observables.

The hydrodynamic flow should yield $\bigl \bra \delta [p_t] \cos(\phi_1^\pm - \phi_2^\pm) \bigl \ket > 0$ in central collisions. The $\vec B$ field effects will add on top of this background. We stress however that, as mentioned earlier in Sec.~\ref{sec:2}, correlations based on \pt{} have so far proven very effective in magnifying small signals. The hope, then, is that, for the correlation of $v_1$ with \pt{}, the contribution due to the genuine $\vec B$ field effects (e.g. CME) will not represent a tiny correction to the hydrodynamic background of order 1\% or lower, but rather of order 5-10\%. This is entirely plausible, and should be investigated in future.

We stress that future theoretical calculations should include as well a more realistic treatment of the structure of the colliding ions, by implementing in particular non-identical proton and neutron densities, as done e.g. in Ref.~\cite{Hammelmann:2019vwd}. The neutron skin may have nontrivial consequences. The edges of the colliding nuclei are mostly populated by neutrons, therefore, the few spectators characterizing a collision at small impact parameter should mostly be neutrons. In a central collision at small \pt{}, like that shown in the left panel of Fig.~\ref{fig:5}, the spectators will all be neutrons. Increasing \pt{}, then, may bring us from a situation where the net $\vec B$ field is literally absent, to one where $\bra B_y \ket >0$. This may eventually lead to a correlation between \pt{} and \By{} that is even stronger than observed in Fig.~\ref{fig:7}.

\section{Acknowledgments}

The work of G.G. is supported by the Deutsche Forschungsgemeinschaft (DFG, German Research Foundation) under Germany’s Excellence Strategy EXC 2181/1 - 390900948 (the Heidelberg STRUCTURES Excellence Cluster), SFB 1225 (ISOQUANT) and FL 736/3-1. C.S is supported by the U.S. Department of Energy (DOE) under grant number DE-SC0013460 and the National Science Foundation (NSF) under grant number PHY-2012922. This work is also supported in part by the U.S. Department of Energy, Office of Science, Office of Nuclear Physics, within the framework of the Beam Energy Scan Theory (BEST) Topical Collaboration.

\end{document}